\documentclass{article}
\usepackage{ijcai19}
\usepackage[english]{babel}

\usepackage{times}
\usepackage{soul}
\usepackage{url}
\usepackage[utf8]{inputenc}
\usepackage[small]{caption}
\usepackage{graphicx}
\usepackage{amsmath}
\usepackage{booktabs}
\usepackage{graphicx}
\title{Automated Detection of COVID-19 from CT Scans Using Convolutional Neural Networks}

\author{
Rohit Lokwani\and
Ashrika Gaikwad\and
Viraj Kulkarni\and
Aniruddha Pant\and
Amit Kharat
\affiliations
DeepTek Inc\\
\emails
}

\begin{document}

\maketitle

\begin{abstract}
COVID-19 is an infectious disease that causes respiratory problems similar to those caused by SARS-CoV (2003). Currently, swab samples are being used for its diagnosis. The most common testing method used is the RT-PCR method, which has high specificity but variable sensitivity. AI-based detection has the capability to overcome this drawback. In this paper, we propose a prospective method wherein we use chest CT scans to diagnose the patients for COVID-19 pneumonia. We use a set of open-source images, available as individual CT slices, and full CT scans from a private Indian Hospital to train our model. We build a 2D segmentation model using the U-Net architecture, which gives the output by marking out the region of infection. Our model achieves a sensitivity of 0.964 (95\% CI: 0.88-1) and a specificity of 0.884 (95\% CI: 0.82-0.94). Additionally, we derive a logic for converting our slice-level predictions to scan-level, which helps us reduce the false positives.

\end{abstract}

\section{Introduction}
Coronaviruses are a large family of RNA viruses which are  usually known to cause respiratory tract illnesses like the common cold. They appear crown-like due to their spiked surface and are categorized into 4 major groups: alpha, beta, gamma, and delta. Most coronaviruses affect animals and can be transmitted between animals and humans \cite{kong2020chest}. COVID-19 is the latest addition to the list of animal-to-human transmissions, preceded by SARS and MERS. COVID-19 is an infectious disease which has affected more than 6.8 million people in the world as of June 8, 2020. The most common clinical manifestations include fever (83\% of patients), cough (82\% of patients), and shortness of breath (31\% of patients) \cite{chen2020epidemiological}. The hallmarks of COVID-19 include bilateral distribution of minute patchy shadows and ground-glass opacity in the nascent stages. The progression of this disease is marked by the spread of these opacities and infiltrates to both the lungs \cite{wang2020deep}.  The World Health Organization has published several testing protocols for detecting the disease \cite{wang2020precision}. The most commonly used reference test for the diagnosis of COVID-19 is the real-time reverse transcription-polymerase chain reaction (RT-PCR) \cite{gundlapallynovel}.

Reverse Transcription Polymerase Chain Reaction (RT-PCR) tests are the key approach used for diagnosing COVID-19. However, they have a few limitations; their shortcomings include the complex process used for specimen collection, the amount of time required for the analysis, and variability in the accuracy of the tests \cite{bullock2020mapping}. Apart from this, a major hurdle in controlling the spread of the disease is the accuracy and shortage of testing kits \cite{zhao2020covid}. Hence, computer-based detection assisted by an expert in the loop with minimal infrastructure is proposed as an alternative to testing kits and vaccines. Computer-aided detection has helped in detecting, localizing, and segmenting out a varied set of diseases using medical imaging analysis. In particular, machine learning is being used for medical imaging analysis by developing deep-learning systems that extract the spatio-temporal representative features from an image, analyze them, and decide the diagnostic outcomes \cite{wang2020deep}.  

The most common, economical, and easy-to-use medical imaging and diagnostic technique is chest radiography or chest X-rays. This technique plays an important role in the diagnosis of lung diseases. Expert radiologists use chest X-ray images (CXRs) to detect pathologies like pneumonia, tuberculosis, atelectasis, infiltrates, and early lung cancer \cite{qin2018computer}. But, detecting COVID-19 using CXRs is challenging due to the less evident visual features in CXRs caused by the overlapping of ribs and soft tissues and low contrast \cite{zhang2020covid}. The limited availability of annotated images adds to the difficulty. The PCR-test is very specific but has a lower sensitivity of 65-95\%, which means that the test can be negative even when the patient is infected \cite{fang2020sensitivity}\cite{ai2020correlation}. These shortcomings can be resolved by using chest CT scans, a cross-sectional imaging modality with high accuracy and speed, instead of CXRs. A recent study of the coronavirus infection on the cruise ship “Diamond Princess” showed evidence of the lung parenchymal pattern (classic for COVID-19) on CT studies of the chest in 54\% of the asymptomatic cases \cite{inui2020chest}.

Most of the recent literature reported that COVID-19-positive patients had characteristic features highly evident in the CT scan images \cite{xie2020chest}. These features included different degrees of ground-glass opacities with or without crazy-paving sign, multifocal organizing pneumonia, and architectural distortion in a peripheral distribution \cite{ai2020correlation}.  COVID-19 eventually develops into chronic pneumonia, and thus the visual symptoms it has are much similar to bacterial and viral pneumonia. In CT scans, the ground-glass opacities are more similar to consolidation \cite{wang2020deep}. Studies have proven that chest CT has a higher sensitivity for the diagnosis of COVID-19 as compared with RT-PCR tests taken from swab samples \cite{ai2020correlation}. To curb human-to-human transmission and isolate the affected from the healthy, it is essential to detect the presence of COVID-19 at an early stage. This is where CT assists in the detection of minor infections \cite{anthimopoulos2016lung}. In this paper, we propose a prospective method using neural networks wherein our model helps in identifying the pixels showing COVID-19 infected regions in a CT scan and helps in marking a patient under consideration as either COVID-19-positive or negative \cite{hesamian2019deep}. 

\section{Related Work}
In the past few years, deep learning has evolved as a technique with its capabilities extending from classification and object detection to segmentation in medical image analysis. Some studies showed better results than expert radiologists.  Rajpurkar et al. \cite{rajpurkar2017chexnet} proposed and presented a DenseNET-121 model for pneumonia detection which performed binary classification on CXRs using CNNs. Qin et al. \cite{qin2018computer} proposed pneumonia and pulmonary edema classification by extracting texture features.  Parveen et al. \cite{parveen2011detection} used an FCM clustering algorithm to detect pneumonia, where they showed that the lung area of the chest was low in black or dark gray when it became infected with pneumonia.

Recently, there have been many developments in detecting COVID-19 from CXRs and CTs. Xu et al. \cite{xu2002deep} proposed a 3-D deep learning model that categorized CTs as either COVID-19 pneumonia-positive or viral pneumonia-positive. They trained a location-attention classification model and used the predicted probabilities to give a prediction calculated by a Bayesian function.  Chen et al. \cite{chen2020deep} built a model using UNet++ \cite{zhou2018unet++}, a powerful architecture for medical image segmentation, and used a 3-consecutive slice and quadrant based post-processing approach to mark a scan as positive or negative. This post-processing approach helped them reduce the number of false positives. Several studies have addressed diagnosis as a binary classification problem, i.e. healthy vs. COVID-19-positive \cite{bullock2020mapping}.  For example, Wang et al. \cite{wang2020deep} used a modified Inception neural network architecture and attained an accuracy of 79.3\%. Szegedy et al. \cite{szegedy2015going} trained on the cropped regions of interest identified by radiologists and distinguished the healthy patients from COVID-19-positive patients.  Several other approaches used a 3-category classification approach, differentiating healthy patients from pneumonia and COVID-19. Xu et al. \cite{xu2002deep} used classical ResNet architectures, adding fully-connected layers at the end, and took the classification approach to solve the problem. He et al. \cite{he2016deep} used ResNets for feature extraction, and Song et al. \cite{song2020deep} used the Feature Pyramid Networks \cite{lin2017feature}, which are the backbone in U-Nets, for learning fine-grained features in the images. Shan et al. \cite{shan2020lung} developed a deep learning system that automatically quantified infection regions of interest (ROIs) and their volumetric ratios with respect to the lung. Li et al. \cite{li2020artificial} put forth a 3D deep learning model, referred to as ConvNet, wherein they combined the 2D local and 3D global features using a max-pooling operation and predicted the class using the probability score from the softmax activation. Jianpeng et al. \cite{zhang2020covid} proposed a deep-learning architecture to differentiate COVID-19 from non-COVID-19 cases from CXRs. Their model is composed of three components: a backbone network, a classification head, and an anomaly detection head. The backbone network extracts the high-level features and feeds it to the rest of the heads. 

Gurujit et al. \cite{randhawa2020machine} identified an intrinsic COVID-19 genomic signature and used it together with a machine learning-based alignment-free approach for an ultra-fast, scalable, and highly accurate classification of whole COVID-19 genomes. 

\section{Data}
We used COVID-19-positive and non-COVID data from GitHub \cite{zhao2020COVID-CT-Dataset} and consolidation and healthy CT scans from a private Indian hospital. The data obtained contained 275 CT scans labeled as COVID-19-positive. The ground truth in these images was decided on the basis of their RT-PCR test results. These CT images had different sizes from 143 patient cases \cite{zhao2020covid}. In total, the data contained 5212 slices and was split into training, validation, and test sets. The prevalence of positive cases in each of the sets was kept at 20\%.  As the available open-source data had varied resolutions, we decided to fix our input size to 512x512 pixels. The original images were in the unsigned int8 format, in the range of [0, 255]. We converted these images to floating-point 16, in the range of [0, 1]. The output masks were in the binary form [0, 1] at pixel-level, where 1s indicated the region of interest. Table 1 shows the detailed distribution of data.

\begin{table}[hbt!]
\begin{center}
\begin{tabular}{ | c | c | c | c |}
\hline
Dataset & COVID-19 & NON-COVID & Total slices\\
\hline
Training & 657 & 2628 & 3285\\
Validation & 120 & 477 & 597\\
Test  & 266 & 1064 & 1330\\

\hline
\end{tabular}
\caption{Slice-level Dataset splits}
\end{center}
\end{table}

The CT slices were annotated, classified, and marked positive by a group of trained expert radiologists. The positive CT slices had typical findings including bilateral pulmonary parenchymal ground-glass and consolidative pulmonary opacities, sometimes with a rounded morphology and a peripheral lung distribution \cite{chung2020ct}. Ground-glass opacification was defined as hazy increased lung attenuation with preservation of bronchial and vascular margins, and consolidation was defined as opacification with obscuration of margins of vessels and airway walls \cite{hansell2008fleischner}. Notably, lung cavitation, discrete pulmonary nodules, pleural effusions, and lymphadenopathy were marked as negative. Our radiologists used an open source tool called VIA \cite{dutta2019vgg} for annotating the images, as it supports various shapes, including ellipses, circles, and polygons, for marking the ROI. figure 1 shows an example of the annotation.

\begin{figure}
\includegraphics[width=\linewidth]{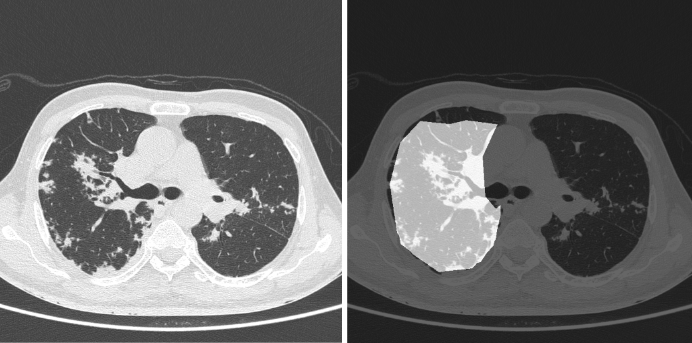}
\caption{(Left) original image and (right) annotated ROI.}
\end{figure}

\section{Methodology}
In this section, we give a brief overview of our training and the inference algorithms.

We used U-Net \cite{ronneberger2015u} for medical image segmentation, which uses the concept of deconvolution \cite{zeiler2014visualizing}. U-Nets are built on the architecture of fully convolutional networks. The most important property of U-Net is the shortcut connections between the layers of equal resolution in the encoder path and the decoder path. These connections provide essential high-resolution features to the deconvolution layers \cite{hesamian2019deep}. Here, we used Xception \cite{chollet2017xception} as the encoder for U-Net.

We used transfer learning by fine-tuning a network pre-trained on CXRs for the same problem but a different task \cite{shin2016deep}. Transfer learning is proven to give better performance when the tasks of source and target network are more similar, and yet even transferring the weights of far and distant tasks has been proved to be better than random initialization \cite{yosinski2014transferable}.

Here, we have tried to solve the problem of distinguishing COVID-19 cases from non-COVID-19 by using weights from our COVID-19 vs Healthy model, as pre-trained weights for this model already gave a sensitivity of 0.9 with a specificity of 0.8. Initially, we built a CT model for consolidation vs healthy and later fine-tuned our model for COVID-19 vs non-COVID-19.

In the training stage, we use binary cross entropy as the loss function and the standard Adaptive Adam Optimizer with a batch size of 4. We set the max epochs to 50 and set the learning rate to $10^{-4}$, which is decayed on the plateau after patience of 4 epochs. We resize each training image to a fixed size of $512 \times 512$ pixels. To alleviate the overfitting of our model on the training data from a particular source, we try to include data from varied sources. One of the drawbacks of having a 2D CT model is that the inference tends to be slow. Our model has a sensitivity of 0.964, hence we plan to use specific slices for inference.

\section{Results}
We tested our model using varied sets of data from different sources. We initially evaluated the model on our test set, consisting of 1330 images, in which COVID-19-positive samples had a prevalence of 20\%. Our model gave a sensitivity of 0.963 (95\% CI: 0.94-0.98) and a specificity of 0.936 (95\% CI: 0.92-0.95). The dice coefficient on positive samples was 0.561. figures 2 and 3 show the superimposed masks on one of the slices.

\begin{figure}
\includegraphics[width=\linewidth]{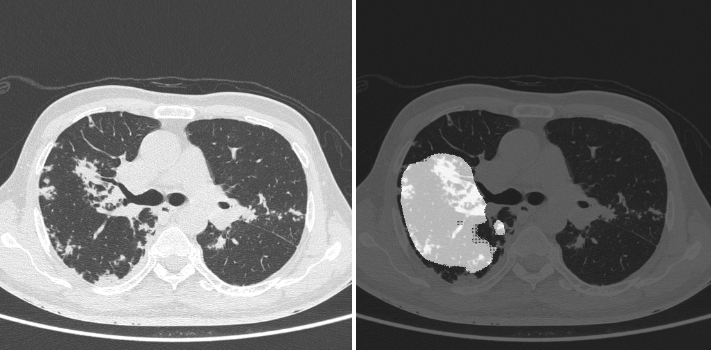}
\caption{(Left) original image and (right) corresponding predicted mask.}
\end{figure}

\begin{figure}
\includegraphics[width=\linewidth]{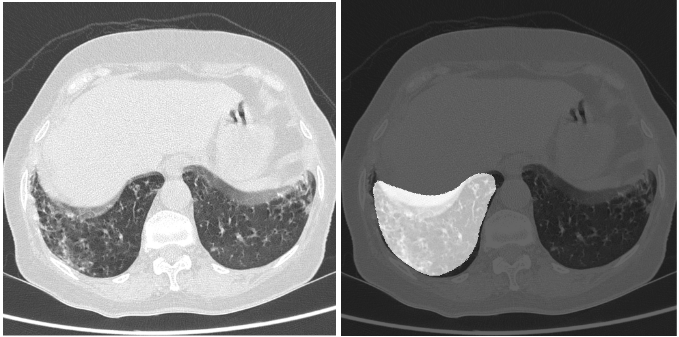}
\caption{(Left) original image and (right) corresponding predicted mask.}
\end{figure}

Apart from this, we evaluated the model on a total of 140 scans with a prevalence of 20\% for positive cases. These scans were tested on data from three sources. One source contained Italian and Chinese scans, while the remaining came from two separate private Indian hospitals. After passing these images through our model, we sorted the slices as per the position of the slice in the CT scan. We observed a pattern wherein the consecutive slices had the same predictions which is expected from a radiology perspective. figure 4 provides an example of the predictions for a positive CT scan. Here we see the expected pattern of consecutive slices, predicted as positive by the model.

\begin{figure}
\includegraphics[width=\linewidth]{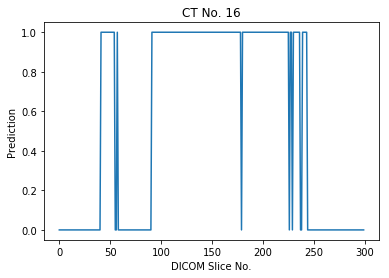}
\caption{Trend of predictions of Slice No. 0 to 300 from CT no. 16.}
\end{figure}

Hence, we convert the slice-level prediction to scan-level prediction using the logic that if 15 consecutive slices in a scan are marked as positive, then we mark the scan as positive. Table 2 shows the results obtained at scan-level.

\begin{table}[hbt!]
\begin{center}
\begin{tabular}{ | c | c | c |}
\hline
Performance Metric & Value & 95\% C.I.\\
\hline
Sensitivity & 0.964 & (0.88,1) \\
Specificity & 0.884 & (0.82,0.94)\\
F1-score & 0.794 & (0.68,0.89) \\

\hline
\end{tabular}
\caption{Scan-level performance of the model on the Test set}
\end{center}
\end{table}

\section{Discussion}
The diagnosis of COVID-19 using CXRs and CT scans has gained significance since the ubiquitous spread of this disease. Given the predominance of ground-glass opacities, chest CT scans usually tend to show the region of infection more clearly than CXRs \cite{kong2020chest}. Our current implementation is a 2D model built at slice-level. Since a CT study could have the number of slices running into thousands, this 2D model certainly adds to the time complexity of processing the whole scan. Although we are satisfied with the performance our model currently shows on the data from diverse distributions, given the time complexity, deploying the model in production is a challenge. In the future, we plan to implement a 3D model that will take the whole CT scan as input and give out masks for the infected areas. The primary challenge with this approach will be the requirement of a lot of annotated data to give a good performance. Additionally, we propose a model that differentiates between COVID-19 and chronic and viral pneumonia and address the  challenges associated with it, like fine-grained, accurate annotations and large amounts of data for all the specified categories. In conclusion, chest CT has proved to have a higher sensitivity than RT-PCR tests \cite{ai2020correlation}. Our analysis suggests that chest CT can be considered for COVID-19 screening and evaluation, especially in epidemic situations where the spread is uncontrollable, and diagnosis needs to be done with celerity. 

\bibliographystyle{ieeetr}
\bibliography{bibliography}
\end{document}